\documentclass[12pt]{article}

\usepackage{amssymb,amsmath}

\DeclareMathOperator{\sech}{sech}
\DeclareMathOperator{\sgn}{sgn}
\DeclareMathOperator{\cn}{cn}

\begin{document}

\begin{titlepage}

\begin{flushright}
arXiv:1902.04643
\end{flushright}
\vskip 2.5cm

\begin{center}
{\Large \bf Statistical Equilibrium of the Korteweg-de Vries and\\
Benjamin-Ono Unidirectional Soliton Models}
\end{center}

\vspace{1ex}

\begin{center}
{\large Brett Altschul\footnote{{\tt baltschu@physics.sc.edu}}}

\vspace{5mm}
{\sl Department of Physics and Astronomy} \\
{\sl University of South Carolina} \\
{\sl Columbia, SC 29208} \\
\end{center}

\vspace{2.5ex}

\medskip

\centerline {\bf Abstract}

\bigskip

The Korteweg-de Vries and
Benjamin-Ono nonlinear wave equations can describe solitary waves,
all of which propagate in the same direction and which
emerge from collisions with their shapes unchanged. This creates a technical challenge to giving a
description of the thermal equilibrium for a gas of such solitons, but the challenges can
be overcome by considering a stochastic source of new solitons located at $x=0$. The total intensity
and the momentum distributions of the solitons emitted by such sources determine the equilibrium
distribution of soliton momenta for $x>0$.

\bigskip

\end{titlepage}

\newpage

\section{Introduction}

Nonlinear field equations that support solitary waves provide explicit field-theoretic models of extended particles,
which propagate without changing their shapes.
Through the study of such solitary waves, it is possible to make qualitative and quantitative determinations about
how extended but
still relatively localized bodies move and interact. These studies have important applications in fluid mechanics,
plasma physics, solid state physics, and elementary particle physics.

Perhaps the most remarkable class of solitary waves are the solitons.
The term {\em solitons} is meant in its oldest and strongest sense, to indicate not merely solitary waves that
can propagate in isolation without changing shapes, but
localized disturbances that pass through one-another and reemerge with their shapes
unchanged. Soliton waveforms will be distorted while they are passing through each another, and although their
original shapes are restored after the collision, their positions in space may be still shifted by the interactions.
This can be interpreted as forward scattering induced by an effective potential between the soliton excitations.
Moreover, some equations that support soliton solutions, such as the sine-Gordon equation $u_{tt}-u_{xx}+\sin u=0$
(using subscript notation for derivatives),
actually have solutions in which the attractive potential between two solitons of different moieties may lead
bound states---``breather'' solutions that expand and contract as the solitons pass back and forth through each
other.

However, the present work will be concerned with a simpler class of soliton-bearing theories. The sine-Gordon
equation describes a relativistic theory for a scalar field in $1+1$ dimensions, but solitons also occur in
nonrelativistic physics, such as in the dynamics of surface waves on a fluid
interface. Another much-studied
$(1+1)$-dimensional nonlinear wave equation is the Korteweg-de Vries (KdV) equation~\cite{ref-boussinesq,ref-korteweg}
\begin{equation}
u_{t}+6uu_{x}+u_{xxx}=0,
\end{equation}
which was discovered---first numerically~\cite{ref-zabusky}, then
analytically~\cite{ref-hirota,ref-gardner1,ref-lax,ref-gardner2}---to support
soliton solutions. The functional form for a single-soliton field profile is easily found by introducing
a traveling ansatz $u=\phi(x-ct)$. The result is
\begin{equation}
\label{eq-kdv-sol}
u(x,t)=\frac{c}{2}\sech^{2}\left[\frac{\sqrt{c}}{2}(x-x_{0}-ct)\right].
\end{equation}
As is usual for a solitary wave, the fixed shape of the
solution represents a balance between the dispersion originating from
the third derivative term in the KdV equation and the equation's nonlinearity.

It is evident from (\ref{eq-kdv-sol}) that any speed $c>0$ is possible, with solitary waves of greater
amplitudes moving more quickly. However, these solitons can only propagate in one direction---the right-moving
direction, toward larger values of $x$.  Unlike the sine-Gordon equation, which is second order in time and thus
invariant under time reversal, the KdV equation and similar equations are only first order in $t$. An 
even simpler equation that supports the propagation of solitons, but only unidirectionally, is the first-order
wave equation $u_{t}+u_{x}=0$, for which any $u=\phi(x-t)$ is a valid solution.

The conceptual framework that we will utilize here---of
a ``gas'' of multiple solitonic excitations of a KdV system---owes a great deal to the work
of Zakharov~\cite{ref-zakharov}, which extracted the behavior of large numbers solitons simultaneously propagating
in parallel. The conception of a soliton gas has proven useful both for analyses of disordered assemblages of many
solitons~\cite{ref-el1}, and for correlated ensembles of solitons forming nearly periodic
lattices~\cite{ref-el2,ref-shurgalina}. At sufficient solitary wave
densities, these systems can have an effectively hydrodynamic
description~\cite{ref-doyon1}. Moreover, 
the statistical methods (using families of random variables~\cite{ref-newell}) used to study wave turbulence
can also be applied to the soliton gas~\cite{ref-el3}.
In such studies of turbulence and other similar contexts, the solutions of the KdV equation can also be used
to construct a nonlinear
filtering method, expanding random functions in KdV eigenmodes instead of Fourier modes~\cite{ref-osborne}.

The KdV equation and other closely related dynamical equations (such as the modified KdV equation,
which has a $u^{2}$ in the nonlinear term~\cite{ref-zhang1}; or the periodic KdV equation, with periodic boundary
conditions limiting long-wavelength behavior)
continue to arise in many different ($1+1$)-dimensional systems---for example, in
magnetized multicomponent plasmas~\cite{ref-samanta,ref-saha}. Because of the ongoing interest in such systems,
analyses of the KdV equation and its relatives remain important areas of research.

Less well known than the KdV equation is the Benjamin-Ono (BO) equation~\cite{ref-benjamin,ref-ono},
\begin{equation}
u_{t}+uu_{x}+H\, u_{xx}=0,
\end{equation}
where $H$ is the Hilbert transform operator, defined by the principal value integral
\begin{equation}
\label{eq-Hilbert}
H\, f(x)=\frac{1}{\pi}{\cal P}\int_{-\infty}^{\infty}d\xi\,\frac{f(\xi)}{x-\xi}.
\end{equation}
The Hilbert transform is clearly nonlocal in coordinate space, but it has a particularly simple representation in
Fourier space, commuting with the spatial derivative $\partial/\partial x$ and
shifting the phases of waves by $\frac{\pi}{2}$, as in
\begin{equation}
H\,\cos(kx+\alpha)=\sgn(k)\cos\left(kx+\alpha-\frac{\pi}{2}\right).
\end{equation}
Note that this phase shift is similar to
the the effect of spatial differentiation, except that it is not accompanied
by multiplicative factor of $-|k|$. The phase shift due to the Hilbert transform is what makes the BO
equation conservative, in contrast to the better-known Burgers' equation~\cite{ref-bateman,ref-burgers}
$u_{t}+uu_{x}-u_{xx}=0$,
which is a nonlinear generalization of the heat equation ($u_{t}-u_{xx}=0$) and thus entails dissipation.

The single-soliton solution of the BO equation is a Lorentzian
\begin{equation}
\label{eq-BO-sol}
u(x,t)=\frac{4c}{1+c^{2}(x-x_{0}-ct)^{2}},
\end{equation}
for $c>0$. The key relations needed to demonstrate that (\ref{eq-BO-sol}) is a solution of the BO equation are
that the Hilbert transform is translation invariant [as is evident from the definition (\ref{eq-Hilbert})],
as well as the specific Hilbert transform
\begin{equation}
\label{eq-H-on-u}
H\,\frac{a}{a^{2}+x^{2}}=\frac{x}{a^{2}+x^{2}}.
\end{equation}
The formula (\ref{eq-H-on-u}), as well as many other Hilbert transforms, are tabulated in Ref.~\cite{ref-poularikas}.

As with the KdV solitons, the field profiles (\ref{eq-BO-sol}) advance more rapidly if they have greater 
amplitudes. However, the character of soliton collisions are different between the two equations.
The KdV solitons emerge from collisions with phase shifts, but the BO solitons do not~\cite{ref-matsuno};
although the BO solitons' profiles are distorted while they overlap in space, their positions after the
interaction are just what they would have been had the interactions not occurred at all. This will make
the statistical analysis of a system of BO solitons relatively simple.

The KdV and BO equations are the simplest nontrivial examples of a large class of unidirectional wave
equations that support soliton solutions. More complicated examples involve additional spatial derivatives
and higher degrees of nonlinearity. As occurs in the BO equation, any term with an even number of derivatives
will be accompanied by a Hilbert transform. (The first-order equation $u_{t}+u_{x}=0$ can be seen as a
degenerate case at the bottom of the whole soliton-bearing equation hierarchy.)
Moreover, aside from their simplicity, the
KdV and BO equations are notable in that they were first studied not for their remarkable mathematical
properties but as models for the behavior of real fluidic systems. The nonlinear $uu_{x}$ terms arise from
the appearance of the convective fluid derivative
$D\vec{u}/Dt=(\partial/\partial{t}+\vec{u}\cdot\vec{\nabla})\vec{u}$---where
$\vec{u}$ is the fluid velocity field---in the Navier-Stokes equation for the momentum.
In contrast, only once the importance of solitons was coming to be appreciated were the higher-order
soliton-supporting equations discovered as theoretical generalizations of the KdV and BO equations.

Of course, the KdV and BO equations, as well as their more elaborate generalizations, are conventionally
expressed in a
partially nondimensionalized form. In a fluid mechanics context the dynamical field $u$ is ultimately 
proportional to the fluid velocity, and for both the KdV and BO solitons, the field profile $u(x,t)$ is
proportional to the soliton speed $c$, times a dimensionless function. However, the
partially nondimensionalized units of the field $u$ and speed $c$ are different between the equations.
In the KdV equation, $u$ and $c$ have units of $($length$)^{-2}$. However,
in the BO equation, the units of these quantities are $($length$)^{-1}$.

One general characteristic of the field equations that support solitonic solutions is that they are
fully integrable and support infinite numbers of conservation laws. (See, for instance,
Refs.~\cite{ref-lax,ref-miura,ref-kruskal} for
the KdV equation and Ref.~\cite{ref-case} for the BO equation.)
That there can be no finite limit
on the number of conserved quantities in these theories is evident from
the following simple, nonrigorous argument. Suppose
that a finite
number of solitons are moving in the $x$-direction. In general, they will all have different amplitudes, and
thus speeds. Given sufficient time, the separations between the solitons may be made arbitrarily large, and
therefore the portion of the total energy that is tied up in the interactions between the solitons may be made
arbitrarily small. Each soliton essentially has a conserved energy associated solely with its own structure and
motion, and since there is no limit on the total number of solitons,
this means that there cannot be any finite limit to the number of conservation laws either.
(In contrast, for a system with
solitary waves that are not true solitons, the solitary waves are never truly isolated from one-another, even at
large physical separations. Interactions between such solitary waves are always accompanied by the emission of
radiation, and even when they are arbitrarily far apart, the solitary waves will be slowly
bleeding away their kinetic energies into finite-wavelength
periodic modes. Thus---and unlike in the soliton case---the energies of such
individual solitary waves are never perfectly conserved, unless there is just a single wave with absolutely
nothing else for it to interact with.) The localized single-soliton energies
are not the usual way in which the infinite families of conservations laws are normally expressed; but
the point of this argument was merely to demonstrate the that existence of an unlimited number of conserved
quantities is intimately tied to the occurrence of soliton behavior.

The object of this paper is to study systems with multiple solitons in thermal equilibrium. When the density of the
solitons is low, they present an example of a gas of particles in one spatial dimension, with exactly solvable
interactions. However, two facts make this analysis somewhat trickier than it might first appear: firstly,
that the solitons are all moving in the same overall rightward direction; and secondly, that the collisions between
solitons never actually alter their speeds. To deal with these complications, it will be useful to consider
solutions of the field equations within a bounded region, with a stochastic source introducing new excitations
at the left-hand end of the region of interest. The equilibrium distribution will then represent a balance between
new solitons being generated by this stochastic source on the left, versus the solitons that are escaping away to
the right. With this formalism, it will be possible to describe the statistical equilibrium of KdV and BO soliton
systems quantitatively.

Moreover, throughout our analysis, it will be worthwhile to keep in mind the fact that nothing about the methodology
depends specifically on the soliton distribution being thermal. The calculations could proceed in the same
manner with a different momentum distribution, such as a power law distribution.
We have selected the thermalized Boltzmann distribution partially
just for the sake of definiteness, although the Boltzmann distribution is special in the sense that it
represents the uniquely likeliest distribution of energies.

However, this work will not be a complete statistical mechanical treatment of either the KdV or BO equation,
because (as alluded to above), in addition to their solitary
waves, both equations support periodic wave train solutions. The periodic solutions of the KdV
equation are known as cnoidal waves, because they involve the Jacobi elliptic function $\cn(z,m)$. Just as for the
solitary waves, the phase speeds of the periodic waves depend on the wave amplitudes, because of the nonlinearity.
In fact, the soliton solutions of both the KdV and BO equations may be found by taking the wavelengths of the
associated periodic solutions to infinity.

The paper is organized as follows. In section~\ref{sec-energies}, we present the energies and momenta of
single-soliton solutions. Section~\ref{sec-BO} then introduces the formalism that will be needed for the analysis of
theories that contain only right-moving solitonic particles. The final results for the BO equation are quite
simple, because it turns out that the BO solitons have a conventional energy-momentum relation,
and because the collisions
between solitons do not actually modify their trajectories. Section~\ref{sec-KdV} applies the same methodology
to a gas of KdV solitons, which is a more complicated problem. The relationship between energy and momentum
is more complicated for a KdV soliton, and pairs of KdV solitons can undergo nontrivial forward scattering
processes. Finally, the conclusions and outlook for future work are summarized in section~\ref{sec-concl}.

\section{Soliton Energies and Momenta}

\label{sec-energies}

For each equation that supports soliton solutions,
there exists a infinite hierarchy of conserved integrals of the motion, involving progressively
higher powers of the field $u$ and its spatial derivatives
(as well as Hilbert transforms in the case of the BO equation and its generalizations).
The two lowest-order conserved quantities represent the total momentum and total energy of the system. These first
two conservation laws also exist for other conservative wave equations, but the higher-order laws that ensure
solitonic behavior are typically absent. Interactions between solitary waves in more general theories
are accompanied by the emission of
radiation; energy is lost from the initial kinetic energies of the solitary waves into propagating wave trains.
Moreover, while individual solitary waves may be topologically protected kinks (interpolating between different vacua),
a collision between a kink and an anti-kink may result in complete annihilation into radiation. Nevertheless, because
other theories that support solitary waves but not true solitons do still typically possess
the first two conserved energy integrals,
the techniques developed in this paper might also be useful for studying
such theories, at least when the density of kinks
(and thus the frequency of collisions) is low.

The energy integrals for the KdV and BO theories are
\begin{eqnarray}
\label{eq-H-KdV}
{\cal H}_{{\rm KdV}} & = & \int dx\left(-\frac{u_{x}^{2}}{2}+u^{3}\right),\\
{\cal H}_{{\rm BO}} & = & \int dx\left[\frac{u\left(H\, u_{x}\right)}{6}+\frac{u^{3}}{18}\right].
\end{eqnarray}
From these, the energies of single-soliton configurations may be determined. (For the KdV case at least,
${\cal H}_{{\rm KdV}}$---treated as a functional of the field $u$---also provides a full Hamiltonian formulation
of the system~\cite{ref-gardner3}.)

There is also a momentum invariant, the same for each equation,
\begin{equation}
P=\int dx\,\frac{u^{2}}{2}.
\end{equation}

For the KdV equation, the energy of the solitary wave may be found by direct integration. The integrand takes the
form
\begin{equation}
-\frac{u_{x}^{2}}{2}+u^{3}=-\frac{c^{3}}{8}\sech^{6}\left(\frac{\sqrt{c}}{2}\xi\right)
\left[\sinh^{2}\left(\frac{\sqrt{c}}{2}\xi\right)-1\right],
\end{equation}
where $\xi=x-x_{0}-ct$. The indefinite integral
\begin{equation}
\int dv\,\sech^{6}v(\sinh^{2}v-1)=-\frac{1}{20}(\cosh 4v+6\cosh 2v+13)\tanh v\sech^{4}v+C,
\end{equation}
along with the asymptotic forms, as $v\rightarrow\pm\infty$,
\begin{eqnarray}
\cosh Nv & \rightarrow & \frac{1}{2}e^{N|v|} \\
\sech^{N}v & \rightarrow & 2^{N}e^{-N|v|} \\
\tanh v & \rightarrow & \sgn v,
\end{eqnarray}
give the energy of the KdV soliton as
$E_{{\rm KdV}}=\frac{1}{5}c^{5/2}$.
Since the overall energy functional ${\cal H}_{{\rm KdV}}$ is really only defined up to an overall multiplicative
constant (related to---among other things---the mass density in a fluid flow problem), the key feature is the
$c^{5/2}$ dependence of the energy on the velocity, which is determined by just the dimensions of the
formula (\ref{eq-H-KdV}) for ${\cal H}_{{\rm KdV}}$.

The integral for the KdV soliton's $P$ is simpler. Using the indefinite integral
\begin{equation}
\int dv\,\sech^{4}v=\frac{1}{3}(\cosh 2v+2)\tanh v\sech^{2}v+C,
\end{equation}
the result is $P_{{\rm KdV}}=\frac{1}{3}c^{3/2}$. Once again, the key fact is that, based on dimensional
considerations, $P_{{\rm KdV}}\propto c^{3/2}$, so that $E_{{\rm KdV}}\propto (P_{{\rm KdV}})^{5/3}$.
The normalizations of the field energy and momentum integrals are not entirely standardized in the literature,
but the normalizations here have been selected so that they obey the correct
velocity relation $c=\partial E/\partial P$ for particle-like excitations.

The same energy and momentum determination procedures may be carried out for the BO equation. For $H\,u_{x}$,
we use the fact that the the Hilbert transform commutes with spatial
differentiation:
\begin{equation}
H\,u_{x}=\frac{\partial}{\partial x}(H\, u);
\end{equation}
combined with the equations (\ref{eq-BO-sol}) and
(\ref{eq-H-on-u}). Then the integrand in the energy $E_{{\rm BO}}$ is
\begin{equation}
\frac{u\left(H\, u_{x}\right)}{6}+\frac{u^{3}}{18}=\frac{(56-24c^{2}x^{2})c^{3}}{9(1+c^{2}x^{2})^{3}},
\end{equation}
and the indefinite integral of this rational function gives
\begin{equation}
\int dv\,\frac{56-24v^{2}}{9(1+v^{2})^{3}}=\frac{2v(19+9v^{2})}{9(1+v^{2})^{2}}+2\tan^{-1}v+C.
\end{equation}
So the result for the energy is $E_{{\rm BO}}=2\pi c^{2}$.
For the momentum $P_{{\rm BO}}$, only the straightforward integral
\begin{equation}
\int dv\,\frac{1}{\left(1+v^{2}\right)^{2}}=\frac{1}{2}\left(\frac{v}{1+v^{2}}+\tan^{-1}v\right)+C
\end{equation}
is required, giving $P_{{\rm BO}}=4\pi c$. Once again, the soliton velocity $c$ is equal to
$\partial E/\partial P$. However, the BO case is also further simplified by that fact that
$E_{{\rm BO}}\propto (P_{{\rm BO}})^{2}$, which is the usual relation for a nonrelativistic particle.

\section{BO Statistics in the Steady State}

\label{sec-BO}

Because of this greater simplicity, and also because the solitons of the BO equation do not experience
any phase delays as a result of passing through each other,
it will be easier to begin the statistical analysis by looking at the BO case.
The theory of the exclusively solitonic BO
excitations is almost equivalent to that of a gas of noninteracting particles in one space dimension.
However, unlike in the standard kinetic theory of gasses, where taking the particles to be (effectively)
noninteracting is an idealization
that simplifies the analysis of the system, in this case, the solitons truly do not interact,
except for the way they distort each other when they are physically overlapping. That actually complicates the
formulation and interpretation of the theory, because the gas will never thermalize on its own. The initial
distribution of soliton energies will remain unchanged in time; there are not even the instantaneous collisions
that reassign momentum and energy in an ideal gas---which quickly move the ideal gas toward the overwhelmingly likely
Maxwell-Boltzmann velocity distribution. A separate complication arises from the fact
that the KdV and BO equations only
support soliton propagation in one direction.

Both of these difficulties can be overcome by considering solitons not along the full $x$-axis, but only
within a finite length $L$. Since the solitons are all moving rightward with different velocities, they will
all eventually leave this finite interval. The solitary waves cannot be reflected, since the BO equation does
not support leftward-moving solutions. Periodic boundary conditions could be imposed, but then the problem that
the soliton energy and momentum distribution functions will never thermalize becomes acute. Instead, we shall
therefore envision the interval of length $L$ being put in contact, at its left end ($x=0$), with a ``thermal''
reservoir of solitons, which enter the interval at random times, traverse the full interval with individual
velocities $c_{i}$, and then leave the interval at $x=L$ after times $L/c_{i}$. In doing this, we will
effectively be treating the solitary waves as pointlike particles. Although this idealization is not
crucial for the BO equation, the assumption that the soliton gas is sufficiently dilute will be
of greater practical importance in our subsequent analysis of the KdV soliton gas.

Understanding the necessary behavior of the thermal source at $x=0$ will be the main goal of this section.
With a satisfactory model of this source, the full behavior of the system follows straightforwardly.

Before proceeding further, a few notation conventions are required.
Since we shall be dealing with densities in the full two-dimensional position-momentum phase space, we
adopt the following terminology. A total extrinsic quantity will be denoted by a capital script letter; the
corresponding spatial density will be denoted by a capital italic letter, as will any other thermodynamically
intrinsic quantity; and the density in momentum
space will be lower case. For example, the total number of solitonic particles ${\cal N}$ contained
in the interval $0\leq x\leq L$ may be expressed as
\begin{equation}
{\cal N}=NL=L\int_{0}^{\infty}dP\, n(P).
\end{equation}

The basic problem will be to determine what conditions must be satisfied in order for the distribution of
solitons in the $0\leq x\leq L$ region to remain thermal. In a situation where the
reservoir at $x=0$ is releasing a thermalized stream of soliton excitations, the momentum distribution of
the solitons must have a Maxwellian form with temperature $T$. Using the BO relation $E=P^{2}/8\pi$ and
normalizing the distribution function appropriately, this yields,
\begin{equation}
n(P)=\frac{N}{\pi\sqrt{2T}}\exp\left(-\frac{P^{2}}{8\pi T}\right).
\label{eq-BO-n}
\end{equation}
Direct calculation of the total internal energy
\begin{equation}
{\cal U}=L\int_{0}^{\infty}dP\, n(P)E(P)=\frac{{\cal N}T}{2},
\end{equation}
produces the expected result that (since $E$ is a quadratic function of $P$) the
average particle energy is $T/2$.

Each soliton remains in the $0\leq x\leq L$ domain only for a time $\tau(P)=L/c(P)=4\pi L/P$, after
which it exits at $x=L$. These departures must be replaced by the stochastic generation of new solitons
at $x=0$ at a rate $S$, where
\begin{equation}
S=\int_{0}^{\infty}dP\, s(P).
\end{equation}
The quantity $s(P)$ is an abstraction of whatever underlying physics is responsible for the thermalization
of the momentum distribution and whatever the solitons are doing before they reach the boundary at
$x=0$.

$\dot{{\cal N}}$, the total average rate of change of the number of solitons, should vanish, so
\begin{eqnarray}
\label{eq-Ndot}
\dot{{\cal N}} & = & \int_{0}^{\infty}dP\,\left[-\frac{n(P)L}{\tau(P)}+s(P)\right] \\
\label{eq-s}
0 & = & \int_{0}^{\infty}dP\,\left[-\frac{1}{4\pi}Pn(P)+s(P)\right].
\end{eqnarray}
In fact, in order that we should not just have $\dot{{\cal N}}$ vanish, but actually
have $\dot{n}=0$, so that
the distribution function is in equilibrium, the integrand in (\ref{eq-s}) must vanish. Thus the
rate at which new solitons are emitted with momenta in the range from $P$ to $P+dP$ is
\begin{equation}
s(P)\,dP=\frac{NP}{4\pi^{2}\sqrt{2T}}\exp\left(-\frac{P^{2}}{8\pi T}\right)\,dP.
\label{eq-BO-s}
\end{equation}
This means that the total emission rate (solitons per unit time) from the thermal source is
\begin{equation}
S=\int_{0}^{\infty}dP\,s(P)=\frac{N}{\pi}\sqrt{\frac{T}{2}}=\frac{N}{\pi}\sqrt{\frac{{\cal U}}{{\cal N}}}.
\end{equation}
Note that because the more energetic solitons spend less time in the $0\leq x\leq L$, they must be emitted
more rapidly than might be guessed just from the equilibrium density distribution $n(P)$. This is the reason
for the extra factor of momentum in (\ref{eq-BO-s}) relative to (\ref{eq-BO-n}). The total emission rate
$S$ also increases with the average energy ${\cal U}/{\cal N}$ for this reason; the more energetic the
mean particle, the faster it reaches $x=L$ and needs to be replaced at $x=0$.

The key ideas of this section can be summarized as follows.
The BO theory can essentially be characterized as a theory for a gas of particles
with only rightward momentum and no collisional interactions at all. The lack of interactions makes things tricky, since
there are no collisions to drive the approach to equilibrium. If the BO equation is satisfied over the entire real line
$-\infty<x<\infty$, then then the density in momentum space $n(P)$ is constant in time, because the momentum
of any given soliton never changes. Thus, in order to grapple with this lack of nontrivial interactions,
it is necessary to introduce a stochastic source. The source distribution $s(P)$ provides the key characterization
of the system; it describes the number of solitons per unit time, per unit momentum, that are emerging into the
$x\geq0$ region. This source of solitons at $x=0$ provides the requisite randomness, and the
system takes on a (one-sided) Maxwellian velocity distribution.
As would be expected for a momentum-dependent quantity arising in a free theory, the source
intensity $s(P)$ is simply proportional to a power of the momentum, times a Boltzmann factor. [If we had used
a different equilibrium distribution $n(P)$, it would have automatically given rise to a different $s(P)$.]

\section{KdV Statistics in the Steady State}

\label{sec-KdV}

The analysis for the KdV solitons will follow along much the same lines as the BO analysis from
section~\ref{sec-BO}. However, the KdV solitons have both a nonstandard energy-momentum relation
and nontrivial forward scattering interactions. As a result, to obtain formulas analogous to all
those obtained in section~\ref{sec-BO} will require additional nontrivial integrations.
There is actually quite a bit of interesting structure to three-body KdV
collisions, while they are going on~\cite{ref-moloney,ref-dimakis,ref-roy}. However, there are no
intrinsic three-body phase shifts~\cite{ref-zakharov}. The asymptotic
position of a soliton after it has passed through two others does not
depend on the relative or absolute positions of those other two; the net shift for the first
soliton is the same whether the other two are widely separated or right on
top of each other at the time of interaction.  This means that the
approximation of a dilute gas is not necessary for the determination of
the average soliton traversal times. There does remain, however, an
implicit assumption that the gas is sufficiently rarefied that it is
possible to speak meaningfully about the locations of individual solitary
waves; for this reason, as well as for computational reasons, we shall treat the interactions perturbatively,
up to ${\cal O}(N^{2})$.

The normalization of the thermal density $n(P)$ with the KdV dispersion relation
$E=(3^{5/3}/5)P^{5/3}$ involves a Gamma function,
\begin{equation}
n(P)=\frac{5^{2/3}N}{\Gamma\left(\frac{3}{5}\right)T^{3/5}}
\exp\left(-\frac{3^{5/3}P^{5/3}}{5T}\right).
\label{eq-KdV-n}
\end{equation}
The numerical constants appearing in this equilibrium density are $3^{5/3}/5\approx 1.248$
and $5^{2/3}/\Gamma(3/5)\approx 1.963$. Using
\begin{equation}
\label{eq-n-int}
\int_{0}^{\infty}dv\,v^{\alpha}e^{-Av^{5/3}}=\frac{3\Gamma\left[\frac{3}{5}(1+\alpha)\right]}
{5A^{\frac{3}{5}(1+\alpha)}},
\end{equation}
the total energy associated with this momentum distribution is
\begin{equation}
{\cal U}=\frac{3{\cal N}T}{5^{11/15}},
\end{equation}
where $3/5^{11/15}\approx0.922$.

The time required for noninteracting solitons to cross $0\leq x\leq L$ was examined in section~\ref{sec-BO}.
In the KdV theory,
the key to determining the required source function $s(P)$ that will maintain the thermal $n(P)$ is
determining how many other solitons a given soliton will collide with during its traversal from
$x=0$ to $x=L$. A soliton will
manage to overtake some of the slower solitons in front of it, and it will
correspondingly be overtaken by some faster-moving ones that start out behind it. The computation of this
effect was not needed for the BO case, because the BO solitons are not displaced during collisions.

When a larger, faster KdV soliton overtakes a smaller, slower one, the faster one emerges from the
interaction with the argument of the $\sech^{2}$ in (\ref{eq-kdv-sol}) advanced by an
amount~\cite{ref-marchant,ref-benes}
\begin{equation}
\label{eq-Delta}
\Delta=\log\frac{\sqrt{c_{1}}+\sqrt{c_{2}}}{\sqrt{c_{1}}-\sqrt{c_{2}}}>0,
\end{equation}
where the solitons have the asymptotic speeds $c_{1}>c_{2}$. The slower soliton emerges with its phase
retarded by an amount $-\Delta$. This effect is often characterized as describing an effective
attractive force between the two bodies. The attraction accelerates the faster, overtaking soliton
and decelerates the slower soliton as the collision approaches. After the collision, the force reverses
directions, and its action eventually returns the two solitons to their original speeds.
The net result is that the faster soliton ends up ahead of where it would have been
in the absence of interaction, while the slower one ends up behind where it
would have been.

The shift $\Delta$ is largest when the two speed are very close together. Collisions of this type are
uncommon, because the relative velocity between the two solitons is small. However,
such collisions, when they do occur, are quite drawn out, giving the effective force between the solitons
a long time to act. This explains why the position shifts are so large in these situations.
Moreover, when $c_{1}$ and $c_{2}$ are close, the actual spatial displacements of the two solitons are
also nearly equal. $\Delta$ is the dimensionless magnitude by which the argument of each soliton's $\sech^{2}$
changes; however, these correspond to different position shifts $\pm2\Delta/\sqrt{c_{i}}$, according to
(\ref{eq-kdv-sol}).

There is also an alternative interpretation, in which the slower soliton leeches energy from the faster one
as the two get close together~\cite{ref-benes}. The amplitude of the front soliton grows,
at the expense of the one behind, until they have interchanged amplitudes. Then the soliton in front speeds
away at its new, faster velocity, and the apparent position shift for the fast soliton just derives from
the fact that the outgoing fast soliton actually started out well in front of the original fast soliton,
before they switched roles.

To determine what $s(P)$ is required to maintain the thermal distribution (\ref{eq-KdV-n}), it is necessary to
know the rate at which solitons reach the right-hand boundary at $x=L$. A soliton
that experienced no interactions would
take a time $L/c(P)$ to cross to $x=L$. However, the presence of other solitons in its path can change the
time required. Treating the average speed for a soliton of momentum $P$ moving through a thermal background
as an unknown function of $P$ produces, in the most general case, an intractable
nonlinear integral equation in momentum space.
However, matters are simplified greatly by approximation that the soliton gas is sufficiently dilute.
Then the predominant contribution to how far a given soliton travels in a time $\tau$ is the free propagation
distance $(3P)^{2/3}\tau$. The distance the the soliton is displaced by any collisions is a small correction on
top of this.

Schematically, we may write that if a soliton (the ``primary'' soliton,
which has momentum $P$) crosses the distance from $0\leq x\leq L$ in
a time $\tau$, we must have
\begin{equation}
L=(3P)^{2/3}\tau+\sum_{P'<P}\frac{2}{(3P)^{1/3}}\log\frac{P^{1/3}+(P')^{1/3}}{P^{1/3}-(P')^{1/3}}
-\sum_{P'>P}\frac{2}{(3P)^{1/3}}\log\frac{P^{1/3}+(P')^{1/3}}{(P')^{1/3}-P^{1/3}}.
\label{eq-distance}
\end{equation}
The initial term on the right-hand-side of (\ref{eq-distance}) is just the free travel distance $c(P)\tau$.
Then the first sum represents a sum over all the solitons that are overtaken as the primary
soliton traverses the interval. Each of these solitons has a momentum $P'<P$, so that they are all moving
more slowly than the primary soliton. If a soliton is less than a distance $L[1-c(P')/c(P)]$ ahead of the
primary soliton when the primary soliton enters the interval at $x=0$, it will be overtaken. So the number
density in momentum space of the solitons that will be overtaken is $L[1-c(P')/c(P)]n(P')$. By an analogous
argument, the momentum space density of $P'>P$ solitons that will overtake the primary soliton is
$L[c(P')/c(P)-1]n(P')$. The position shifts due to these are collected in the second sum
on the right-hand side of (\ref{eq-distance}).
Note that both the $P'<P$ and $P'>P$ densities are positive.

The summands in each of the sums are just $2\Delta/\sqrt{c(P)}$. These are the forward or backward displacements
corresponding to the phase differences $\Delta$, according to (\ref{eq-kdv-sol}) and (\ref{eq-Delta}).

The $P'<P$ sum in (\ref{eq-distance}) is therefore
\begin{equation}
\sum_{P'<P}\Big(\cdots\Big)=\frac{2L}{(3P)^{1/3}}\int_{0}^{P}dP'
\left\{\left[1-\left(\frac{P'}{P}\right)^{1/3}\right]n(P')
\log\frac{P^{1/3}+(P')^{1/3}}{P^{1/3}-(P')^{1/3}}\right\},
\end{equation}
while the $P'>P$ sum is correspondingly
\begin{equation}
-\sum_{P'>P}\Big(\cdots\Big)=-\frac{2L}{(3P)^{1/3}}\int_{P}^{\infty}dP'
\left\{\left[\left(\frac{P'}{P}\right)^{1/3}-1\right]n(P')
\log\frac{P^{1/3}+(P')^{1/3}}{(P')^{1/3}-P^{1/3}}\right\}.
\end{equation}
These two integrals can evidently be combined to give a single expression,
\begin{equation}
\label{eq-KdV-tau}
\tau=\frac{L}{(3P)^{2/3}}-\frac{2L}{3P}\left\{\frac{5^{2/3}N}{\Gamma\left(\frac{3}{5}\right)T^{3/5}}
\int_{0}^{\infty}dP'\left[1-\left(\frac{P'}{P}\right)^{1/3}\right]\exp\left[-\frac{3^{5/3}(P')^{5/3}}{5T}\right]
\log\left|\frac{P^{1/3}+(P')^{1/3}}{P^{1/3}-(P')^{1/3}}\right|
\right\},
\end{equation}
using the $n(P')$ from (\ref{eq-KdV-n}).

Following the same reasoning as with (\ref{eq-Ndot}) in section~\ref{sec-BO}, it follows that the source of
solitons at $x=0$ must be emitting them with a rate density in momentum space of
\begin{equation}
s(P)=\frac{n(P)L}{\tau(P)}.
\end{equation}
Treating the collision tern in (\ref{eq-KdV-tau}) as a small correction, this involves
\begin{equation}
\frac{L}{\tau(P)}=(3P)^{2/3}\left[1+\frac{2(15^{2/3})}{\Gamma\left(\frac{3}{5}\right)}
\frac{NP^{2/3}}{T^{3/5}}f\left(P^{5/3}/T\right)\right],
\end{equation}
where the function $f(u)$ contains the complicated integral,
\begin{equation}
f(u)=\int_{0}^{\infty}dv\,\left[v^{2}(1-v)\exp\left(-\frac{3^{5/3}}{5}v^{5}u\right)\log\left|
\frac{1+v}{1-v}\right|\right]
\end{equation}

For low temperatures (meaning large $u$), the integral for $f(u)$ is tractable, since it will be dominated by
the small-$v$ integration region. Using
\begin{equation}
\int_{0}^{\infty}dv\,v^{3}e^{-Av^{5}}=\frac{\Gamma\left(\frac{4}{5}\right)}{5A^{4/5}},
\end{equation}
the source strength in this limit is
\begin{equation}
\label{eq-large-T}
s(P)=\frac{15^{2/3}NP^{2/3}}{\Gamma\left(\frac{3}{5}\right)T^{3/5}}\exp\left(-\frac{3^{5/3}P^{5/3}}{5T}\right)
\left[1+\frac{4(5^{7/15})\Gamma\left(\frac{4}{5}\right)}{3^{2/3}\Gamma\left(\frac{3}{5}\right)}\frac{NT^{1/5}}
{P^{2/3}}\right].
\end{equation}
The additional constants appearing in this expression are $15^{2/3}/\Gamma(3/5)\approx4.084$ and 
$4(5^{7/15})\Gamma(4/5)/3^{2/3}\Gamma(3/5)\approx3.186$.

It makes sense that, when the temperature is sufficiently small compared with the scales set by the soliton density
$N$ and the primary soliton momentum $P$, the collision term is small. The reason lies in the formula
(\ref{eq-Delta}) for $\Delta$. When a fast-moving, energetic soliton overtakes a much slower one, the displacements
caused by $\Delta$ are small, because the argument of the logarithm in (\ref{eq-Delta}) is nearly unity. In short,
solitons with wildly different momenta barely affect each other, because they pass through each other so quickly.

If the ${\cal O}(N^{2})$ collision term is entirely neglected, we may relate the source strength $S$ to the total
energy ${\cal U}$, as in section~\ref{sec-BO}. For such free flowing solitons, using (\ref{eq-n-int}) again,
\begin{equation}
S=\frac{5^{2/3}}{\Gamma\left(\frac{3}{5}\right)}NT^{2/5}=\frac{5^{24/25}N}{3^{2/5}\Gamma\left(\frac{3}{5}\right)}
\left(\frac{{\cal U}}{{\cal N}}\right)^{2/5}.
\end{equation}
The power of ${\cal U}/{\cal N}$ involved
is determined by dimensional considerations alone, but the overall constant
$5^{24/25}/3^{2/5}\Gamma(3/5)\approx2.203$ is nontrivial.

Ultimately, we see that the statistical KdV theory possesses all the complications of the analogous BO theory,
plus some additional ones.
Even though the BO soliton gas represents an essentially free theory, the fact that the solitons only move
unidirectionally complicates affairs, and in order to have steady-state thermal behavior, it is necessary to
have an input of new solitons entering the
region of interest from the left. This feature obviously continues to be important when the solitons are
governed by the KdV equation. Moreover, the solitons in the KdV theory do have nontrivial pairwise interactions. All
collisions are completely elastic, with no momentum redistribution. However, the solitons are attracted toward
one-another, and each two-body collision alters the positions of the solitons involved. In this section, we have
generalized the statistically modeling techniques used to study the BO system to address the KdV equation,
with its
nonstandard energy-momentum relation and nontrivial interactions. To get tractable results, the interactions
were treated perturbatively, up to second order in the soliton density $N$. In order to get a closed form
expression (\ref{eq-large-T}) without the complicated integral function $f(u)$, it was also necessary to look
at the limit of high temperatures.

\section{Conclusions and Outlook}

\label{sec-concl}

We have seen how to construct a theory of statistical equilibrium for gasses of KdV and BO solitons. These two
theories share a number of complications; in particular, they only support right-moving solitary waves, and the
solitary waves are true solitons---meaning that there is no redistribution of momentum during collisions between
two or more solitons.  For the solitons of the BO theory, these are essentially the only complications of note.
The BO solitons have a conventional dispersion relation, with $E_{{\rm BO}}\propto(P_{{\rm BO}})^{2}$, just like
for ordinary nonrelativistic particles. The KdV system, in contrast, is more subtle.

Throughout our analysis, we have
assumed the existence of thermalized Boltzmann equilibrium distributions. However, as previously noted,
it would be entirely possible to carry through the same kind of analysis with a
momentum distribution function $n(P)$ that had a different functional form.
(In fact, some detailed simulations of the periodic KdV equation have indeed used entirely
different soliton energy distributions~\cite{ref-dutyhk}.)
In practice, our calculations took the thermal distribution of
soliton momenta as an assumption and worked backwards from there to determine the necessary source function
$s(P)$. Starting with a different $n(P)$, it would be possible to get a different $s(P)$.  This raises the
natural question of whether a $s(P)$ function like (\ref{eq-BO-s}) or (\ref{eq-large-T})
is really special in any way. In fact, these particular
forms for $s(P)$ do have special significance,
precisely  because they gives rise to a thermal distributions of soliton momenta. When the
number of solitons is thermodynamically large, the
conventionally structured thermal distribution is unique in that it represents the most likely way of dividing
up the total energy among the many solitonic particles.
Perturbations to the KdV and similar equation have been shown to have a thermalizing
effect~\cite{ref-malomed1,ref-malomed2}. If the interactions between solitons are perturbed by
weak soliton-soliton momentum exchange interactions, the state of the gas will evolve toward a thermal
distribution. In our model, the soliton source at $x=0$ is a ``black box''; it may be that for $x<0$,
the field equation governing the solitons is slightly modified, so that they
do interact in that domain, and the soliton gas in the $x<0$ region thermalizes before the solitons are injected into
the $0\leq x\leq L$ region of interest.

An alternative viewpoint would be to start with a known distribution of injected solitons $s(P)$ and work
in the other direction---deriving the equilibrium $n(P)$ that corresponds to such a $s(P)$. This is a
straightforward generalization of the approach described here. However, other generalizations of our approach may
be trickier to implement. For instance, the fact that a number of simplifying approximations
were used in section~\ref{sec-KdV} raises the obvious possibility that it
might be possible to further improve the results---to have closed form expressions that would be valid
beyond the small $N$, small $T$ limit. An even more ambitious, and perhaps more important, generalization
would be to consider a statistical treatment that includes the full spectrum of solutions of the BO or KdV
equations. These nonlinear wave equations have periodic wave train solutions, with the soliton solutions
representing only the infinite wavelength limits. Nontrivial interactions between finite-wavelength waves will
make matters significantly more complicated, in part because there is no notion of the excitations being far apart.
A further adjustment to how the thermal theory is formulated may be necessary in order to obtain a tractable
description.

However, even with just the soliton results presented in this paper, progress has been made, both conceptually
and computationally, toward a
complete theory of statistical equilibrium for fully integrable systems. The resulting
picture of thermal equilibrium may be useful for characterizing plasmas,
turbulently interacting surface waves, or other
systems with weakly interacting nonlinear dynamics.

\end{document}